\documentclass{article}

\usepackage{arxiv}

\usepackage[utf8]{inputenc} 
\usepackage[T1]{fontenc}    
\usepackage{hyperref}       
\usepackage{url}            
\usepackage{booktabs}       
\usepackage{amsfonts}       
\usepackage{nicefrac}       
\usepackage{microtype}      
\usepackage{lipsum}
\usepackage{graphicx}
\graphicspath{ {./images/} }

\usepackage{balance} 
\usepackage{listings}
\usepackage{relsize}

\title{\textbf{raison}: Developing Reliable Decision-Making Agents}

\author{
 Pavlos Moraitis \\
  Argument Theory\\
  Paris, FR \\
  Université Paris Cité\\
  Paris, FR \\
  \texttt{pavlos.moraitis@u-paris.fr} \\
   \And
 Nikolaos Spanoudakis \\
  Argument Theory\\
  Paris, FR \\
  Hellenic Mediterranean University\\
  Chania, GR \\
  \texttt{nspan@hmu.gr} \\
  \And
 Antonis Kakas \\
  Argument Theory\\
  Paris, FR \\
  University of Cyprus\\
  Nicosia, CY\\
  \texttt{antonis@ucy.ac.cy} \\
}

\begin{document}
\maketitle
\begin{abstract}
This paper presents the \textbf{\textit{r{\smaller[2.2]AI}son}} platform, a high-level technological environment for the development of automated, reliable and explainable decision-making agents. The research underlying the platform and its technological progress has now reached a mature stage that allows the platform to be used for the development of complex real-life applications without writing a single line of code. 
\end{abstract}
\section{The rAIson Platform}
The recent developments in AI have given rise to a renewed interest in building autonomous agents. In this effort, one of the main challenges is related to the reliability of the decisions and actions of the agents.
The platform of \textit{r{\smaller[2]AI}son} addresses this challenge by providing an environment for supporting the development of reliable and explainable, decision making agents. It has recently been developed by the \href{https://www.argument-theory.com/}{``Argument Theory''} start-up company. 

The wider aim of \textit{r{\smaller[2.2]AI}son} is to enable individuals to innovate in their own field by providing them with the capability of developing decision-making systems directly from their high-level understanding of their  expertise without writing a single line of code. The developers and users of the \textit{r{\smaller[2]AI}son} platform need not be aware of the underlying technology of the platform, much in the same way as in using Large Language Models (LLMs). In comparison to LLMs, in \textit{r{\smaller[2]AI}son} we concentrate on building systems under the human's expertise of the application problem offering \textit{absolute reliability} with respect to the reasoning capability of the developed systems. This reliability is guarantied by the use of \textbf{Gorgias }~\cite{AAMAS-KM03,GorgiasAppArg2019,DBLP:conf/comma/SpanoudakisGKK22}, a generic inference engine that is endowed with both \textit{argumentative} and \textit{abductive} reasoning capabilities. This system implements the \textit{structured preference-based argumentation framework }called \textbf{Logic Programming with Priorities } \cite{AAMAS-KM03}.
The core technology of \textbf{\textit{r{\smaller[2]AI}son}} consists of two components:
\begin{itemize} 
    \item A Front-end \textit{authoring tool} for modeling complex decision-making problems through \textit{natural language dialogue,} directly in the application domain. Developers can construct decision policies—i.e., decision and preference arguments expressing the relative strength of one argument over another—simply by answering the platform’s questions in natural language, without writing any code. This dialogue-driven approach automates \textit{knowledge acquisition} and \textit{preference elicitation}, enabling the scalable development of symbolic AI systems.
    \item A Back-end with explainable argumentation as a decision-making engine, using the Gorgias system, offered \textit{as a Service} over the Internet \textbf{(AIaaS)}. The explanations are derived from the knowledge supplied by the developer throughout the knowledge acquisition and preference elicitation process. 
\end{itemize}

These components are complemented by services for \textit{automated code }and \textit{API generation}, as well as \textit{deployment.} The \textit{r{\smaller[2]AI}son} platform complemented with these services (tools) supports the development of automated decision-making agents by assisting  developers in the specification, implementation, deployment and testing of their systems. Moreover, being offered as a service, this decision making capability can be interfaced with other ``standard'' agent platforms, e.g. the Java Agent Development Framework (JADE)~\cite{bergenti2020first}, or other technologies, e.g. Internet of Things (IoT)~\cite{spanoudakis2024iot}, to build agentic application systems.   

The authoring tool of the platform is based on the \textbf{Software Development via Argumentation (SoDA)} methodology for knowledge engineering and acquisition~\cite{DBLP:conf/ecai/SpanoudakisKM16,GorgiasAppArg2019,DBLP:conf/icaart/SpanoudakisKM21}. In this methodology, knowledge, e.g., a Decision Policy, is captured in terms of \textbf{hierarchies of Scenario Based Preferences (SBP)}, namely statements expressing which options are preferred under an application scenario and how such preferences change in successively more specific scenarios~\cite{GorgiasAppArg2019}. 
An \textit{option} can be a decision, an action, a belief or a goal. Consider, for example, the following part of a Decision Policy for a Salary Negotiation Agent: "\textit{If the offered salary is above the expected salary, accept the offer. Refuse an offer when the offered salary is low. Also refuse an offer when the offered salary is close to the expected salary, but accept it if the yearly salary increase plan, brings within two years the salary well above the expected salary}."

Decision arguments in \textbf{\textit{r{\smaller[2]AI}son}} have the form  \textbf{$\mathit{Scenario \rhd Option}$}. In the above decision policy, $Options=\{accept, refuse\}$ are the two possible decisions while $Scenarios$=\{the offered salary is above the expected salary, the offered
salary is low, the offered salary is close to the expected salary, the yearly salary increase plan brings within two years the salary well above the expected salary\} are the possible scenarios supporting the options (i.e. the decisions) and the preferences.

By responding to the natural-language questions posed by \textit{r{\smaller[2]AI}son} through an advanced authoring tool, developers can define options and their associated scenarios to construct decision arguments. By applying a \textit{conflict resolution process}, they can then establish hierarchies of preference arguments—expressing unconditional or conditional (contextual) preferences—to resolve conflicts between competing decision or preference arguments at different levels (i.e. meta-preferences). Once this process is completed, developers can verify the resulting decision policies and trigger (with a \textit{run} button) their automatic translation into Gorgias language by \textit{r{\smaller[2]AI}son} (see Listings \ref{fig:NL} and \ref{fig:FOL}).
The generated code is immediately automatically deployed on the platform, making the developed system ready for testing and use—either through a run screen in \textit{r{\smaller[2]AI}son} or via the API. In other words, the authoring tool of the \textit{r{\smaller[2]AI}son}  platform offers a \textit{``no-code programming''} environment for the software development of the decision-making modules of agents.
Developers can work in two modes: \textit{basic} and \textit{advanced}. In basic mode, they use only natural language to represent options and argument (decision and preference) scenarios. This representation is automatically translated by \textit{r{\smaller[2]AI}son} into propositional Gorgias code (see Listing \ref{fig:NL}). 

\begin{lstlisting}[caption=Basic mode Gorgias code.,breaklines=true,showstringspaces=false,tabsize=1,label={fig:NL}]
rule(r1,accept,[]):-the_salary_offered_is_above_the_expected_salary.
rule(r2,refuse,[]):-the_offered_salary_is_low.
rule(r3,refuse,[]):-the_offered_salary_is_close_to_the_expected_salary.
rule(r4,accept,[]):-the_offered_salary_is_close_to_the_expected_salary.
rule(p1,prefer(r3,r4),[]).
rule(p2,prefer(r4,r3),[]):-the_yearly_salary_increase_plan_brings_within_two_ years_the_salary_well_above_the_expected_salary.
rule(c1,prefer(p2,p1),[]).
complement(accept,refuse).
complement(refuse,accept).
\end{lstlisting}


This level of representation is often sufficient for domains based on qualitative knowledge, such as legal, medical or regulatory compliance contexts. In advanced mode, developers can combine natural language with 
\href{https://www.swi-prolog.org}{Prolog},
enabling them to refine (or personalize) the automated representation using first-order logic (FOL)—introducing variables, predicates, functions, or mathematical expressions. Based on this new formal representation, \textit{r{\smaller[2.2]AI}son} automatically generates a new code (see Listing \ref{fig:FOL}, which personalizes Listing \ref{fig:NL}). This is particularly useful in domains where formal precision is required, such as finance, trading, risk evaluation, or negotiations (i.e. our example). This dual-mode approach supports collaboration between users with different roles and technical expertise. For example, a human resources manager may define recruitment decision policies in natural language (Listing \ref{fig:NL} for our example), while a technical expert formalizes them using first-order logic (Listing \ref{fig:FOL} for our example). The FOL representation is done through a dedicated GUI.  \textbf{Both automatically generated codes are completely transparent to the developers (basic or advanced).} Interested readers can view videos showcasing the development of various use-cases and tutorials available here: \textit{www.youtube.com/@Argument-Theory.}

Decision-making is then carried out via argumentation-based reasoning of building arguments for options that are acceptable, i.e., that are stronger or at least as strong, with respect to preference arguments expressing preferences for arguments supporting contrary options (see rules r1 and r2) or contrary preferences (see rules p1 (unconditional preference) and p2 (conditional/contextual preference)) at different hierarchical levels. Rule c1 expresses unconditional meta-preference.

By examining the corresponding rules (i.e., those with the same names) in Listings \ref{fig:NL} and \ref{fig:FOL}, one can observe how expressions such as " the offered salary is above the expected salary", “the offered salary is low", ”the offered salary is close to the expected salary” or “the yearly salary increase plan brings
within two years the salary well above the expected salary” are formally represented. Alternative formal encodings, however, are also possible. 
\textit{Complements} represent the conflict between competing options. \textit{Brackets} in the rules allow to declare \textit{beliefs} (defeasible knowledge) or \textit{abducibles} (hypotheses) when applicable.

\begin{lstlisting}[caption=Advanced mode Gorgias code.,breaklines=true,showstringspaces=false,tabsize=1,label={fig:FOL}]
rule(r1,accept,[]):-offered_salary(O),expected_salary(E),O>=E.
rule(r2,refuse,[]):-offered_salary(O),expected_salary(E),O=<0.7*E.
rule(r3,refuse,[]):-offered_salary(O),expected_salary(E),E>O, O>0.7*E.
rule(r4,accept,[]):-offered_salary(O),expected_salary(E),E>O, O>0.7*E.
rule(p1,prefer(r3,r4),[]).
rule(p2,prefer(r4,r3),[]):-offered_salary(O),expected_salary(E),
    yearly_salary_increase(X),O*((1+X)**2)>1.5*E.
rule(c1,prefer(p2,p1),[]).
complement(accept,refuse).
complement(refuse,accept).
\end{lstlisting}


One main advantage of using an argumentation-based computation is the fact that this naturally gives informative \textbf{explainable solutions} \cite{DBLP:conf/comma/SpanoudakisKK22,DBLP:journals/ws/SpanoudakisGKK23,ATheodorou2024}. 
Gorgias can explain its reasoning process by tracing the decision and preference arguments it employed to reach a decision, as well as the data on which the decision was based.



To bring the \textit{r{\smaller[2]AI}son} decision-making services into the practice of building real world agents, the platform exposes an \textbf{API} with two major services:
a) \textbf{\textit{Get application metadata}}: this service is of type GET and returns the ids of the elements that make up the possible scenarios and the ids of the options
b) \textbf{\textit{Query an application}}: this service is of type POST and allows the agent to query the developed application for allowed options within a given context composed of scenario elements.
These services allow the integration of a decision policy to an agent decision-making module. 

The \textit{Gorgias} technology has been applied in a wide variety of real-life applications (see e.g. \cite{GAID2024}, 
\cite{Sofia2022}, 
\cite{MEDICA2017}, 
\cite{Erisa2020},
\cite{marcais2011using},
\cite{moraitis2007argumentation}).
Some recent applications are built using the \textit{r{\smaller[2]AI}son} platform 
(\cite{mandikas2025effective},
\cite{Michalakis2025}). 
These applications have been developed using the \textit{r{\smaller[2]AI}son} under a closed, regulated access. 
Since last summer, the \textit{r{\smaller[2]AI}son} platform is also open for beta testing.
The \textit{r{\smaller[2]AI}son} platform currently supports decision problems involving several dozen options, several hundred decision rules, and several thousand lines of generated Gorgias code overall (including preference and meta-preference rules).
Recently, we initiated research work on a \textit{neuro-symbolic approach} aimed at enhancing the platform’s efficiency and usability, when human expertise is not available, through the introduction of novel features.
\textit{r{\smaller[2]AI}son} will open to a global paying audience through Amazon Marketplace in the coming months.


\bibliographystyle{unsrt}  
\bibliography{references}  

@inproceedings{mandikas2025effective,
  title={An Effective Root-Finding Toolbox Using Computational Argumentation},
  author={Mandikas, Vasileios G and Spanoudakis, Nikolaos I},
  booktitle={2025 6th International Conference in Electronic Engineering \& Information Technology (EEITE)},
  year={2025},
  doi={10.1109/EEITE65381.2025.11166218},
  organization={IEEE}
}

@inproceedings{marcais2011using,
  title={Using argumentation for ambient assisted living},
  author={Marcais, Julien and Spanoudakis, Nikolaos and Moraitis, Pavlos},
  booktitle={International Conference on Engineering Applications of Neural Networks},
  pages={410--419},
  year={2011},
  doi={10.1007/978-3-642-23960-1\_48},
  organization={Springer}
}

@article{moraitis2007argumentation,
  title={Argumentation-based agent interaction in an ambient-intelligence context},
  author={Moraitis, Pavlos and Spanoudakis, Nikolaos},
  journal={IEEE Intelligent Systems},
  volume={22},
  number={6},
  pages={84--93},
  year={2007},
  doi={10.1109/MIS.2007.101},
  publisher={IEEE}
}

@inproceedings{spanoudakis2024iot,
  author={Spanoudakis, Nikolaos I. and Krinakis, Panteleimon and Kolokotsa, Dionysia},
  title={A Methodology for Applying Decision Policies into Smart Buildings with the use of Computational Argumentation and IoT Technologies},
  booktitle={5th International Conference in Electronic Engineering, Information Technology \& Education (EEITE 2024)},
  year={2024},
  doi={10.1109/EEITE61750.2024.10654451},
  organization={IEEE}
}

@article{bergenti2020first,
  title={The first twenty years of agent-based software development with JADE},
  author={Bergenti, Federico and Caire, Giovanni and Monica, Stefania and Poggi, Agostino},
  journal={Autonomous Agents and Multi-Agent Systems},
  volume={34},
  number={2},
  pages={36},
  year={2020},
  publisher={Springer},
  doi={10.1007/s10458-020-09460-z}
}

@article{GAID2024,
AUTHOR = {Tanos, Panayiotis and Yiangou, Ioannis and Prokopiou, Giorgos and Kakas, Antonis and Tanos, Vasilios},
TITLE = {Gynaecological Artificial Intelligence Diagnostics (GAID) GAID and Its Performance as a Tool for the Specialist Doctor},
JOURNAL = {Healthcare},
VOLUME = {12},
YEAR = {2024},
NUMBER = {2},
ARTICLE-NUMBER = {223},
URL = {https://www.mdpi.com/2227-9032/12/2/223},
PubMedID = {38255110},
ISSN = {2227-9032},
DOI = {10.3390/healthcare12020223}
}

@inproceedings{DBLP:conf/comma/SpanoudakisKK22,
  author       = {Nikolaos Spanoudakis and
                  Antonis C. Kakas and
                  Adamos Koumi},
  editor       = {Kristijonas Cyras and
                  Timotheus Kampik and
                  Oana Cocarascu and
                  Antonio Rago},
  title        = {Application Level Explanations for Argumentation-based Decision Making},
  booktitle    = {1st International Workshop on Argumentation for eXplainable {AI} co-located
                  with 9th International Conference on Computational Models of Argument
                  {(COMMA} 2022), Cardiff, Wales, September 12, 2022},
  series       = {{CEUR} Workshop Proceedings},
  volume       = {3209},
  publisher    = {CEUR-WS.org},
  year         = {2022}
}

@article{DBLP:journals/ws/SpanoudakisGKK23,
  author       = {Nikolaos I. Spanoudakis and
                  Georgios Gligoris and
                  Adamos Koumi and
                  Antonis C. Kakas},
  title        = {Explainable argumentation as a service},
journal = {Journal of Web Semantics},
volume = {76},
pages = {100772},
year = {2023},
doi = {10.1016/j.websem.2023.100772},
}

@inproceedings{DBLP:conf/comma/SpanoudakisGKK22,
  author       = {Nikolaos I. Spanoudakis and
                  Georgios Gligoris and
                  Antonis C. Kakas and
                  Adamos Koumi},
  editor       = {Francesca Toni and
                  Sylwia Polberg and
                  Richard Booth and
                  Martin Caminada and
                  Hiroyuki Kido},
  title        = {Gorgias Cloud: On-line Explainable Argumentation},
  booktitle    = {Computational Models of Argument - Proceedings of {COMMA} 2022, Cardiff,
                  Wales, UK, 14-16 September 2022},
  series       = {Frontiers in Artificial Intelligence and Applications},
  volume       = {353},
  pages        = {371--372},
  publisher    = {{IOS} Press},
  year         = {2022},
  doi={10.3233/FAIA220178}
}

@inproceedings{DBLP:conf/icaart/SpanoudakisKM21,
  author       = {Nikolaos I. Spanoudakis and
                  Konstantinos Kostis and
                  Katerina Mania},
  editor       = {Ana Paula Rocha and
                  Luc Steels and
                  H. Jaap van den Herik},
  title        = {Web-Gorgias-B: Argumentation for All},
  booktitle    = {Proceedings of the 13th International Conference on Agents and Artificial
                  Intelligence, {ICAART} 2021, Volume 2, Online Streaming, February
                  4-6, 2021},
  pages        = {286--297},
  publisher    = {{SCITEPRESS}},
  year         = {2021},
  doi={10.5220/0010269402860297}
}

@inproceedings{DBLP:conf/ecai/SpanoudakisKM16,
  author       = {Nikolaos I. Spanoudakis and
                  Antonis C. Kakas and
                  Pavlos Moraitis},
  editor       = {Gal A. Kaminka and
                  Maria Fox and
                  Paolo Bouquet and
                  Eyke H{\"{u}}llermeier and
                  Virginia Dignum and
                  Frank Dignum and
                  Frank van Harmelen},
  title        = {Applications of Argumentation: The SoDA Methodology},
  booktitle    = {{ECAI} 2016 - 22nd European Conference on Artificial Intelligence,
                  29 August-2 September 2016, The Hague, The Netherlands},
  series       = {Frontiers in Artificial Intelligence and Applications},
  volume       = {285},
  pages        = {1722--1723},
  publisher    = {{IOS} Press},
  year         = {2016},
  doi={10.3233/978-1-61499-672-9-1722}
}

@article{GorgiasAppArg2019,
  author       = {Antonis C. Kakas and
                  Pavlos Moraitis and
                  Nikolaos I. Spanoudakis},
  title        = {\emph{GORGIAS}: Applying argumentation},
  journal      = {Argument \& Computation},
  volume       = {10},
  number       = {1},
  pages        = {55--81},
  year         = {2019},
  doi={10.3233/AAC-181006}
}

@article{Sofia2022,
author = {Almpani, Sofia and Kiouvrekis, Yiannis and Stefaneas, Petros and Frangos, Panayiotis},
title = {Computational Argumentation for Medical Device Regulatory Classification},
journal = {International Journal on Artificial Intelligence Tools},
volume = {31},
number = {01},
pages = {2250005},
year = {2022},
doi={10.1142/S0218213022500051}
}

@article{Erisa2020,
title = {An Argumentation-Based Reasoner to Assist Digital Investigation and Attribution of Cyber-Attacks},
journal = {Forensic Science International: Digital Investigation},
volume = {32},
pages = {300925},
year = {2020},
author = {Erisa Karafili and Linna Wang and Emil C. Lupu},
doi={10.1016/j.fsidi.2020.300925}
}

@inproceedings{MEDICA2017,
  author       = {Nikolaos I. Spanoudakis and
                  Elena Constantinou and
                  Adamos Koumi and
                  Antonis C. Kakas},
  editor       = {Salem Benferhat and
                  Karim Tabia and
                  Moonis Ali},
  title        = {Modeling Data Access Legislation with Gorgias},
  booktitle    = {Advances in Artificial Intelligence: From Theory to Practice - 30th
                  International Conference on Industrial Engineering and Other Applications
                  of Applied Intelligent Systems, {IEA/AIE} 2017, Arras, France, June
                  27-30, 2017, Proceedings, Part {II}},
  series       = {Lecture Notes in Computer Science},
  volume       = {10351},
  pages        = {317--327},
  publisher    = {Springer},
  year         = {2017},
  doi={10.1007/978-3-319-60045-1\_34}
}

@InProceedings{ATheodorou2024,
author="Methnani, Leila
and Dignum, Virginia
and Theodorou, Andreas",
title="Clash of the Explainers: Argumentation for Context-Appropriate Explanations",
booktitle="Artificial Intelligence. ECAI 2023 International Workshops",
year="2024",
publisher="Springer Nature Switzerland",
address="Cham",
pages="7--23",
doi={10.1007/978-3-031-50396-2\_1}
}

@inproceedings{AAMAS-KM03,
  author    = {Antonis C. Kakas and
               Pavlos Moraitis},
  title     = {Argumentation based decision making for autonomous agents},
  booktitle = {Proc. of 2nd Int. Joint Conf. on Autonomous Agents {\&}
               Multiagent Systems, {AAMAS}},
  pages     = {883--890},
  year      = {2003},
  publisher = {{ACM}},
  doi={10.1145/860575.860717}
}

@mastersthesis{Michalakis2025,
    author      = {Ioannis Michalakis},
    title       = {Argumentation-based decision support system for systems deployment: case study in the Ministry of Digital Governance},
    type        = {Diploma Thesis}, 
    school = {Technical University of Crete},
    year        = {2025},
    doi={10.26233/heallink.tuc.104026}}

\end{document}